\documentclass[12pt]{article}
\usepackage{a4}
\usepackage[dvips]{graphicx}
\usepackage{amssymb}
\usepackage{mathbbol}
\begin{document}
\begin{titlepage}
\vskip0.5cm
\begin{flushright}
DESY 02-200\\
\end{flushright}
\vskip0.5cm
\begin{center}
{\Large\bf Speeding up 
 Lattice QCD simulations with clover-improved Wilson Fermions}
\vskip 0.3cm
{\Large\bf }
\vskip 0.3cm
\end{center}
\vskip 1.3cm
\centerline{
M. Hasenbusch and K. Jansen
}
\vskip 0.4cm
\centerline{\sl  NIC/DESY-Zeuthen}
\centerline{\sl  Platanenallee 6, D-15738 Zeuthen, Germany}
\vskip 0.3cm
\centerline{e-mail: Martin.Hasenbusch@desy.de, Karl.Jansen@desy.de}

\vskip 0.4cm
\begin{abstract}
We apply a recent proposal to speed up the Hybrid-Monte-Carlo simulation
of systems with dynamical fermions to two flavour QCD with 
clover-improvement.
The basic idea of our proposal is to split 
the fermion matrix into two factors with a reduced condition number each.
In the effective action, for both factors a pseudo-fermion field 
is introduced.
For our smallest quark masses we see a speed-up of more
than a factor of two compared with the standard algorithm. 
\end{abstract}
\end{titlepage}

\section{Introduction}
It is clear that with simulation algorithms that are used today
it will be very difficult, if not impossible,
to reach the physical values of lightest quark masses. The scaling 
behaviour for Wilson fermions (see \cite{Lippert,Ukawa,Wittig,panel2001,ECFA}) 
of the algorithms predicts enormous costs for simulations 
at quark masses as light as the u- and d-quarks. To reach this 
physical point, extrapolations using chiral perturbation theory ($\chi$PT)
have to be used. However contact to $\chi$PT seems to be happening at 
rather small values of the quark masses themselves (see ref. \cite{panel2002}
and refs. therein).
Any progress to render simulations easier, when approaching the small
quark mass regime will help therefore to reach the overlap region between 
$\chi$PT and lattice QCD allowing for a safe extrapolation to
physical quark masses.

Still the HMC (Hybrid-Monte-Carlo) algorithm \cite{hybrid}
and its variants \cite{forcrand,Jansenrev,PeardonLattice} are the 
methods of choice
in simulating lattice QCD with dynamical Wilson fermions.
Recent large scale simulations with two flavours of Wilson fermions  
and standard boundary conditions
\cite{CP-PACS,Alltonetall}
only reach a ratio of $m_{\pi}/m_{\rho} \gtrapprox 0.58$, while the physical 
point is given by $m_{\pi}/m_{\rho} \approx 0.18$. 
Note that the decay
of the $\rho$-meson into two $\pi$-mesons is only possible if 
$m_{\pi}/m_{\rho}<0.5$. 
In refs. \cite{Irving,CP-PACS2002} 
explorative studies with $m_{\pi}/m_{\rho} \gtrapprox 0.4$ were reported and
it was found that the simulations become substantially more
expensive.                   
Going to light quarks, the HMC algorithm becomes increasingly 
expensive for at least
two reasons:  First the condition number of the fermion matrix increases.
Hence the number of iterations needed by the solver employed 
increases. Secondly, but maybe related, the 
step-size of the integration scheme has to be decreased to maintain a 
constant acceptance rate.  In fact, in ref. \cite{Irving} a step size as
small as $1/400$ is needed for the leap-frog integration scheme.

In ref. \cite{MH_schwinger} we have demonstrated that the obstacle
of very small step sizes at low values of the quark mass
can be lessened by a modification of the pseudo-fermion action.
The proposal is to  split the fermion matrix into two
factors and to introduce a pseudo-fermion field for both factors.
Each of the corresponding two matrices 
has a smaller condition number than the original fermion matrix.
As a consequence the ``fermion force'', i.e. the
response of the system on a variation of the gauge field, is
substantially reduced.
The numerical study of the two dimensional Schwinger model showed that
the step-size can be enlarged and thus the
computational effort can be reduced substantially this way.
In ref. \cite{ourlattice2001} we presented first results for lattice
QCD with two flavours of Wilson fermions and clover improvement 
\cite{clover}.
In the present paper we extend this study towards larger lattices and 
smaller quark masses. Also, we compare two different factorisations
of the fermion matrix. 
We simulated at $\beta=5.2$ on $8^3 \times 24$ and  $16^3 \times 24$ lattices.
To allow a comparison with the literature, we have taken the values for
the parameters $c_{sw}$ and $\kappa$ from ref. \cite{Sroczynski}.
Our largest value of $\kappa$ corresponds  to $m_{\pi}/m_{\rho}=0.698$.
Results of simulations with Schr\"odinger functional boundary 
conditions are reported in ref. \cite{Knechtlietal}. A brief summary of 
the present work can be found in \cite{ourlattice2002}.

The paper is organised as follows: In section 2 we briefly recall the
definition of clover-improved Wilson fermions 
\cite{clover,alphacsw2,alphacsw3}. As basic
improvement we employ even-odd preconditioning as discussed in \cite{JansenLiu}.
Next we discuss in detail the modified pseudo-fermion action 
\cite{MH_schwinger} and a variant of it \cite{ourlattice2001}. We explain
how (easily) the HMC algorithm can be adopted to the modified pseudo-fermion 
action. It follows a thorough discussion of the integration schemes that we
have used. In section 4 we give the details of our simulation. Next 
we present our numerical results and discuss its implications.
Finally we give our conclusions and an outlook.

\section{The modified pseudo-fermion action}

In this section we remind the reader of the action 
of clover-improved Wilson fermions. For completeness, 
we briefly recall even-odd 
preconditioning and the standard form of the pseudo-fermion 
action  that is used in the HMC simulation.
Then we show how the modified pseudo-fermion action that was 
proposed in ref. \cite{MH_schwinger}
can be generalised to clover-improved fermions. In addition to the 
original proposal, we consider an alternative that is inspired 
\cite{Sommerprivate} by twisted mass QCD \cite{twisted}
and was first presented in 
ref. \cite{ourlattice2001}. Finally, we explain how the HMC algorithm can 
be adopted to the modified pseudo-fermion action.

\subsection{The model}
Our aim is to simulate the system defined by the partition function
\begin{equation}
\label{partition}
 Z = \int \mbox{D}[U] \exp(-S_G[U])  \;\; \mbox{det} M[U]^2 \;\;,
\end{equation}
where the Wilson plaquette action is given by
\begin{equation}
S_G[U] = - \frac{\beta}{3} \sum_{x} \sum_{\mu>\nu} 
\mbox{Re} \; \mbox{Tr} \; 
\left ( U_{x,\mu} U_{x+\hat \mu,\nu} U_{x+\hat \nu,\mu}^{\dag} 
U_{x,\nu}^{\dag} \right ) \;\;,
\end{equation}
where
$x$ are sites on a hyper-cubical lattice, 
$\mu,\nu \in \{0,1,2,3\}$ are directions
on the lattice and $\hat \mu$ is a unit vector in $\mu$-direction.
In eq.~(\ref{partition}), the
fermion degrees of freedom have been integrated out. For 
Wilson fermions with clover-improvement, the fermion matrix $M$ is given 
by \cite{clover}
\begin{eqnarray}
 M[U]_{xy} &=& 
 \left(1 - \frac{i}{2} \; c_{sw} \; \kappa \; \sigma_{\mu \nu} \; 
  {\cal F}_{\mu \nu}(x) \right) \delta_{x,y} \nonumber \\
&-& \kappa \sum_{\mu} \left\{ 
 (1-\gamma_{\mu} ) \; U_{\mu}(x) \; \delta_{x+\hat \mu,y} +
 (1+ \gamma_{\mu} ) \; U_{\mu}^{\dag}(x-\hat \mu)\; \delta_{x-\hat \mu,y}
\right\} \;\; ,
\end{eqnarray}
where we sum over $\mu$ and $\nu$. 
The anti-symmetric and anti-Hermitian tensor ${\cal F}$ is given by
\begin{eqnarray}
 {\cal F}_{\mu,\nu} &=& \frac{1}{8} 
   \left [U_{\mu}(x) U_{\nu}(x+\hat \mu) U_{\mu}^{\dag}(x+\hat \nu)
    U_{\nu}^{\dag}(x) \right. \nonumber \\
 &+&U_{\nu}(x) 
    U_{\mu}^{\dag}(x+\hat \nu -\hat \mu) 
    U_{\nu}^{\dag}(x -\hat \mu)
    U_{\mu}(x-\hat \mu) \nonumber \\
 &+&U_{\mu}^{\dag}(x-\hat \mu) 
    U_{\nu}^{\dag}(x-\hat \nu -\hat \mu) 
    U_{\mu}(x-\hat \nu -\hat \mu)
    U_{\nu}(x-\hat \nu) \nonumber \\
 &+&U_{\nu}^{\dag}(x-\hat \nu) 
    U_{\mu}(x-\hat \nu) 
    U_{\nu}(x-\hat \nu +\hat \mu)
    U_{\mu}^{\dag}(x) \nonumber \\
 &-& \left. h.c. \right]  \;\;\;.
\end{eqnarray}
For a discussion of on-shell O(a)-improvement see e.g. refs.
\cite{clover,alphacsw2,alphacsw3,JansenLiu,alphacsw1}.

In our simulations we have used even-odd preconditioning throughout.
We followed the proposal of ref. \cite{JansenLiu}, see also \cite{luo}.
The fermion matrix can be written as 
\begin{equation}
 M=  \left( \begin{array}{cc}
    \mathbb{1}_{ee}+T_{ee}   &  -\kappa M_{eo}  \\
              -\kappa M_{oe}   &  \mathbb{1}_{oo}+T_{oo}
\end{array} \right)  \; \; \;,
\end{equation}
where we have introduced the matrix $T_{ee}$($T_{oo}$) on the
even (odd) sites as
\begin{equation} \label{eq:t}
(T)_{xa\alpha,yb\beta} =
{i \over 2}c_{sw}\kappa\sigma^{\alpha\beta}_{\mu\nu}
{\cal F}^{ab}_{\mu\nu}(x) \delta_{xy}\;\;.
\end{equation}
The off-diagonal parts $M_{eo}$ and $M_{oe}$, which
connect the even with odd and odd with even lattice
sites, respectively, are just
the conventional Wilson hopping matrices.
The determinant of the fermion matrix can now be written as
\begin{equation}
\label{ourevenodd}
\mbox{det} M \propto \mbox{det} (\mathbb{1}_{ee} + T_{ee}) \; \mbox{det} \hat M
\;\;,
\end{equation}
where
\begin{equation}
\hat M = 
\mathbb{1}_{oo} + T_{oo} - M_{oe} (\mathbb{1}_{ee} + T_{ee})^{-1} M_{eo}
\;\;.
\end{equation}
In the following discussion we shall refer to the Hermitian matrix
\begin{equation}
\label{hatQ} \label{c0hat}
 \hat Q = \hat c_0 \gamma_5 \hat M
\end{equation}
with $\hat c_0 $ a constant set to $\hat c_0 = 1$ throughout this work.

In the standard HMC simulation of mass-degenerate two-flavour Wilson fermions,
the effective action
\begin{equation}
\label{standardaction}
 S_{eff}[U,\phi^{\dag},\phi] = 
 S_G[U] + S_{det} [U] + S_F[U,\phi^{\dag},\phi] \;\;,
\end{equation}
with 
\begin{eqnarray} \label{eq:actionsf}
  S_{det}[U] &  = & -2 \mbox{Tr} \log(1+T_{ee}) \nonumber \\
  S_F[U,\phi^{\dag},\phi] &  = & \phi^{\dag} \hat Q^{-2} \phi
\end{eqnarray}
is used.
We shall refer to $S_{det}[U]$ as the determinant contribution. The 
pseudo-fermion action $S_F[U,\phi^{\dag},\phi]$ is based on the representation
\begin{equation}
\label{standardpseudofermions}
\mbox{det} \hat Q^2 \propto \int \mbox{D} \phi^{\dag} \int \mbox{D} \phi
\exp(-\phi^{\dag} \hat Q^{-2} \phi)
\end{equation}
of the square of the fermion determinant. In our study we keep $S_G[U]$ 
and $S_{det}[U]$ in their standard form.  
However, 
$S_F[U,\phi^{\dag},\phi]$ is replaced by modified expressions to be 
discussed below.

As an alternative, the authors of ref. \cite{JansenLiu} suggest a more 
symmetrical treatment of even and odd sites. The result is  
\begin{equation}
\label{symm}
\mbox{det} M \propto \mbox{det} (\mathbb{1}_{ee} + T_{ee})
                     \mbox{det} (\mathbb{1}_{oo} + T_{oo}) 
                     \mbox{det} \hat M_{sym} \;\;\;,
\end{equation}
where now 
\begin{equation}
\hat M_{sym} = \mathbb{1}_{oo}  - 
         (\mathbb{1}_{oo} + T_{oo})^{-1} 
         M_{oe} (\mathbb{1}_{ee} + T_{ee})^{-1} M_{eo} \;\;\;.
\end{equation}
The authors of ref. \cite{JLQCD} find that the choice of eq.~(\ref{symm}) leads 
to a roughly $30 \%$ higher performance of the HMC algorithm than  
the choice of eq.~(\ref{ourevenodd}). 

Since we wanted to compare our results with those of ref. \cite{Sroczynski}
and we started with our study before ref. \cite{JLQCD} appeared, we
have used only eq.~(\ref{ourevenodd}) in this study.
However, it is straight-forward to apply our modification to eq.~(\ref{symm})
and it can be expected that a similar gain should be found.

\subsection{The modified pseudo-fermion action}
Our starting point of modifying the action of eq.~(\ref{eq:actionsf})  
is the observation that, at fixed acceptance rate and length
of the trajectory, the step-size of the integration scheme has to be decreased
with decreasing sea-quark masses and hence with increasing condition number 
of the fermion matrix. This effect can be nicely seen e.g.
in table II of ref. \cite{CP-PACS}.

Also, a number of studies (see e.g. \cite{Pe,alphabench,FoTa96})  
have shown that  replacing the original fermion matrix by a preconditioned
fermion matrix in the pseudo-fermion action allows for a larger 
step-size in the HMC simulation at the same acceptance rate.

Based on these observations, one of us \cite{MH_schwinger} proposed to
factorise the fermion matrix into two parts, with reduced
condition number each. 
The determinant of both factors is estimated by
pseudo-fermion fields:
\footnote{In refs. \cite{alphabench,MHlocal} we applied a similar modification 
of the pseudo-fermion action to speed up the local updating of lattice QCD.}
\begin{eqnarray}
\label{general}
  \mbox{det} \hat Q^2  &=&  
\mbox{det} W W^{\dag}\; \mbox{det} [W^{-1} \hat Q]
[W^{-1} \hat Q]^{\dag}
\;\;\; \propto \;\;\; 
 \int \mbox{D} \phi_1^{\dag} \int \mbox{D} \phi_1 \;
 \int \mbox{D} \phi_2^{\dag} \int \mbox{D} \phi_2 \nonumber \\
& &
 \phantom{xxx} 
\exp\left(-\phi_1^{\dag} \left(W W^{\dag}\right)^{-1} \phi_1
-\phi_2^{\dag} \left([W^{-1} \hat Q] 
               [W^{-1} \hat Q]^{\dag} \right)^{-1} \phi_2
\right) \;\;. 
\end{eqnarray}
Note that $W$ might be non-Hermitian.
In the following we shall use the notation
\begin{equation}
S_{F1} = 
\phi_1^{\dag} \left(W W^{\dag}\right)^{-1} \phi_1 \;\;, \;\;\;\;
S_{F2} =  \phi_2^{\dag} \left([W^{-1} \hat Q] 
         [W^{-1} \hat Q]^{\dag}\right)^{-1} \phi_2 \;\;\;.
\end{equation}
In ref.~\cite{MH_schwinger} we also considered even-odd preconditioning,
but no clover-improvement. We constructed $W$ by a shift in the 
hopping-parameter $\kappa$: $W = \gamma_5 \tilde M$, where
\begin{equation}
\tilde M=\mathbb{1} - \tilde \kappa^2  M_{oe} M_{eo} \;\; ,
\end{equation}
with {\em $\tilde \kappa < \kappa$}. Since the fermion matrix can be multiplied
by a constant factor without changing the physics, we can write equivalently
\begin{equation}
\label{originaltilde} 
W = \hat Q + \rho \gamma_5 \;\;.
\end{equation}
This is the form that we apply to the even-odd preconditioned, clover-improved 
 $\hat Q$ of eq.~(\ref{hatQ}).
In ref. \cite{MH_schwinger} we have demonstrated at the example of the 
two-dimensional Schwinger model that the step-size of the integration scheme
can indeed be increased compared with the standard pseudo-fermion action. This 
gain increases as the sea-quark mass decreases. For the lightest mass that 
we studied, the step-size could be increased by more than a factor of two
for the optimal choice of $\tilde \kappa$.

In addition to the original choice (\ref{originaltilde}) of $W$ 
we shall consider
\begin{equation}
\label{rainertilde}
\tilde{W} = \hat Q + i \rho \mathbb{1} \;\;,
\end{equation}
which is inspired \cite{Sommerprivate} by twisted mass QCD \cite{twisted} and
was first tested in ref. \cite{ourlattice2001}.

In terms of $\hat Q$ the resulting pseudo-fermion action is 
$S_F=S_{F1}+S_{F2}$ with
\begin{equation}
\label{orginal}
 S_{F1} = \phi_1^{\dag} [\hat Q + \rho \gamma_5]^{-2} \phi_1 \;\;\;\; , \;\;\;
 S_{F2} = \phi_2^{\dag}
            [\mathbb{1} + \rho \gamma_5 \hat Q^{-1}] 
            [\mathbb{1} + \rho \hat Q^{-1} \gamma_5] 
    \phi_2 
\end{equation}
for eq.~(\ref{originaltilde}) and
$\tilde{S}_F=\tilde{S}_{F1}+\tilde{S}_{F2}$ with
\begin{equation}
\label{rainer}
 \tilde{S}_{F1} = \phi_1^{\dag} [\hat Q^{2} + \rho^2 \mathbb{1}]^{-1} \phi_1
 \;\;\;\; , \;\;\;
 \tilde{S}_{F2} = \phi_2^{\dag} [\mathbb{1} +\rho^2 \hat Q^{-2} ] \phi_2 
\end{equation}
for eq.~(\ref{rainertilde}).

The aim of our modification is to use 
matrices with reduced condition number in the pseudo-fermion action.
Let us denote the smallest and the largest eigenvalue of $\hat{Q}^2$ by
$\lambda_{min}$ and $\lambda_{max}$, respectively.
If we  choose $\lambda_{min} \ll \rho^2  \ll \lambda_{max}$, 
we find the condition number of $\hat Q^2+\rho^2$ to be $\lambda_{max}/\rho^2$.
On the other hand, the condition number of the second matrix used 
in $\tilde{S}_{F2}$ of eq.~(\ref{rainer}) 
becomes $\rho^2/\lambda_{min}$. This suggests an optimal choice
of  \footnote{We are grateful to R. Sommer for this argument.}
\begin{equation}
\label{ruleforrho}
 \rho^2 = \sqrt{ \lambda_{max} \lambda_{min} } \;\;.
\end{equation}
In the simulations with the pseudo-fermions using the action $\tilde{S}_F$
of eq.~(\ref{rainer}), as described below, we followed this rule. 
The obvious disadvantage of this choice
is that the eigenvalues $\lambda_{min}$ and $\lambda_{max}$ have to be computed
to a reasonable accuracy before the simulation with the modified pseudo-fermions
can be done with optimised choices of the algorithm. 
Also it is not clear,  how much the detailed distribution 
of the eigenvalues of $\hat Q^2$ matters for the step size of 
the integration scheme used.

Therefore we did not use the
 rule of eq.~(\ref{ruleforrho}) in the simulations with
the action of eq.~(\ref{orginal}).  Instead, we tried to find numerically the 
value of $\rho$, where for fixed step-size the acceptance rate is the largest.

As a last remark in this section we want to mention that with the choice 
of $\rho^2$ of eq.~(\ref{ruleforrho}) it seems that the condition number
$k$ of the original action, eq.~(11), is reduced to $\sqrt{k}$ when 
using the action given in eq.~(\ref{rainer}).
 If hence the scaling behaviour of the 
HMC algorithm is indeed determined to a large extend by the condition 
number of the fermion matrix used, a much better scaling behaviour 
can be expected when the chiral limit is approached. 

\subsection{The Hybrid-Monte-Carlo Algorithm}
\label{HMC}
For completeness, we recall the steps of the Hybrid-Monte-Carlo algorithm
\cite{hybrid}
applied to the standard pseudo-fermion action of eq.~(\ref{standardaction}).
One elementary update (``trajectory'') of the HMC algorithm is composed
of the following steps:
\begin{itemize}
\item Global heat-bath of the pseudo-fermions and the conjugate momenta.
\item Molecular dynamics evolution of the gauge-field and the conjugate
      momenta $P$ with fixed pseudo-fermions.
\item Accept/Reject step: the gauge-field $U'$ that is generated by the 
      molecular dynamics evolution is accepted with the probability \\
      $P_{acc}= \mbox{min} [1, \exp(-H(U',P',\phi)+H(U,P,\phi))]$,\\
      where $P'$ represents the conjugate momenta generated in the 
      molecular dynamics evolution.
\end{itemize}

The last step is needed, as 
the molecular dynamics evolution can not be done exactly and hence
requires a numerical integration scheme, 
rendering $\Delta H = H(U,P,\phi) - H(U',P',\phi) \ne 0$.
To obtain a valid algorithm, 
this scheme has to be area preserving and reversible. Details on the 
numerical integration schemes that we have studied 
are given in section \ref{schemes}.

\subsection{Heat-bath of the pseudo-fermion fields}
The field $\phi_1$ is initialised at the beginning of the trajectory
as usual\footnote{Here we discuss only the case of the action 
$S_F$ in eq.~(\ref{orginal}). The case of the action $\tilde{S}_F$ in 
eq.~(\ref{rainer}) is basically identical.}:
\begin{equation}
 \phi_1= W \eta_1 \;\;,
\end{equation}
where $\eta_1$ has a Gaussian distribution.
The heat-bath of $\phi_2$ however requires the application of the inverse
of $W$:
\begin{equation}
\label{heat2}
\phi_2= W^{-1} \hat Q \eta_2 \;\;,
\end{equation}
where also $\eta_2$ has a Gaussian distribution.

\subsection{The variation of the pseudo-fermion action}
\label{trajsect}
An important feature of our approach is that 
the variation of the modified pseudo-fermion action can be computed as easily 
as in the standard case.
For the first part of the pseudo-fermion action we obtain
upon a variation of the action with respect to the gauge fields
\begin{equation}
\label{firstdiff}
\delta S_{F1} = - X^{\dag} \; \delta W\;   Y  \;\;
 - Y^{\dag} \; \delta W^{\dag} \; X\;
\end{equation}
with the vectors
\begin{equation}
\label{XYgeneral1}
 X = \left(W W^{\dag} \right)^{-1} \; \phi_1  \;\;,
  \;\; Y = W^{-1} \; \phi_1 \;\;\;,
\end{equation}
which is essentially the same as for the standard pseudo-fermion action.
The only difference is that $\hat Q$ is replaced by $W$.

For the second part of the  pseudo-fermion action we get
\begin{equation}
\label{seconddiff}
\delta S_{F2} = - X^{\dag} \; \delta \hat Q \;   Y  \;\;
 - Y^{\dag} \; \delta \hat Q^{\dag} \; X
\;\; + X^{\dag} \; \delta W \; \phi_2 \;\;
 + \phi_2^{\dag} \; \delta W^{\dag} \; X
\end{equation}
with the vectors
\begin{equation}
\label{XYgeneral2}
 X = \left(\hat Q \right)^{-2} \;W \; \phi_2  \;\;,
 \;\; Y = \hat Q^{-1} \;W \; \phi_2 \;\;\;.
\end{equation}

For our choices of $W$
in eq.~(\ref{originaltilde}) (and of $\tilde{W}$ in eq.~(\ref{rainertilde}))
the variation of $S_{F2}$ (and $\tilde S_{F2}$)
further simplifies and it holds $\delta W = \delta \hat Q$ ($=\delta \tilde{W}$).
We therefore get
\begin{equation}
\delta S_{F2} = - X^{\dag} \; \delta \hat Q \;   Y  \;\; 
 - Y^{\dag} \; \delta \hat Q \; X\; \;\;\;, 
\end{equation}
where
\begin{equation}
\label{XYorginal2}
 X =  \hat Q^{-1} [\mathbb{1} + \rho \hat Q^{-1} \gamma_5] \; \phi_2  \;\;,\;\;\;
 Y =  \rho \hat Q^{-1} \gamma_5 \; \phi_2 \;\;\;
\end{equation}
for  the action in eq.~(\ref{orginal}) 
and
\begin{equation}
\label{XYrainer2}
 X = \rho^2 \hat Q^{-2} \; \phi_2  \;\; , \;\;\; Y = \hat Q^{-1} \; \phi_2 \;\;
\end{equation}
for the action in eq.~(\ref{rainer}).

Let us discuss the effect of introducing the parameter $\rho^2$ on
the variation of the pseudo-fermion actions. 
Since the step
size used in the HMC algorithm increases as the condition number 
decreases, we expect that we can choose a larger step size 
for the action $\tilde S_{F1}$. For the action $\tilde S_{F2}$ we see from 
eq.~(\ref{XYrainer2}) that its variation includes a factor of
$\rho^2$. Since $\rho^2 \ll 1$, 
we expect again smaller variations of the action along a trajectory
leading 
to larger choices of the step size. A similar argument holds for the action
$S_F$ of eq.~(\ref{orginal}).

\section{Integration schemes}
\label{schemes}
We have tested two different integration schemes: The standard leap-frog
and a partially improved one suggested by Sexton and Weingarten 
(see eq.~(6.4) of ref. \cite{SeWe})
that still has an  $\delta \tau^2$ error as the leap-frog scheme.
However the amplitude of this  error is considerably reduced.

Let us define the update of the gauge-field and the momenta as
\begin{eqnarray}
 T_U (\delta \tau) \; &:& \; U \rightarrow e^{i \delta \tau \; P }
                         \; U \\
 T_P (\delta \tau) \; &:& \;  P \rightarrow P 
  - i \delta \tau \;
                      \delta_U (S_g(U) + S_f(U))  \;\;\;,
\end{eqnarray}
where $\delta_U$ denotes a variation with respect to the gauge fields.

The elementary step of the leap-frog algorithm is given by
\begin{equation}
\label{ourleap}
 T_2(\delta \tau) = T_P \left(\frac{\delta \tau}{2} \right) \; T_U(\delta \tau) \;
                    T_P \left(\frac{\delta \tau}{2} \right) \;\;\;.
\end{equation}
A trajectory is composed of $N_{md}$ 
consecutive  elementary steps.
Here we use trajectories of length 1, i.e. $N_{md} \; \delta \tau=1$, as
it is done in most HMC simulations.
Note that the order of the updates of momenta and gauge-fields is not unique.
In fact, in ref. \cite{JLQCD} it was demonstrated that the alternative order
\begin{equation}
\label{altleap}
 T_2'(\delta \tau) = T_U \left(\frac{\delta \tau}{2} \right) \; T_P(\delta \tau) \;
                    T_U \left(\frac{\delta \tau}{2} \right) \;
\end{equation}
achieves the same acceptance rate as eq.~(\ref{ourleap})
(see fig. 2 of ref. \cite{JLQCD}) with  a roughly $15 \%$ larger step size
$\delta \tau$.

Since we like to compare our numerical results with those of ref. 
\cite{Sroczynski} we have used the order of eq.~(\ref{ourleap}) in our numerical
study. Also it is not clear to us how the idea of a split time-scale
introduced in ref. \cite{SeWe} can be applied to eq.~(\ref{altleap}).

In the literature \cite{SeWe,CrGo,CaRo} 
also higher-order schemes are discussed. These schemes
become increasingly complicated as the order increases. Recent studies 
\cite{CP-PACS,Takaishi,UKQCDInstable}
show that higher-order schemes become less efficient than the 
simple leap-frog integration scheme of eq.~(\ref{ourleap})
as $\kappa_c$ is approached.

In addition to the genuinely higher-order schemes, Sexton and Weingarten \cite{SeWe}
discuss a scheme with reduced $\delta \tau^2$ errors. In the numerical 
tests, using Wilson fermions, 
on a $4^4$ lattice at $\beta=5.4$ and $\kappa=0.162$, 
Sexton and Weingarten  \cite{SeWe} found that this scheme 
achieves the best performance. 
The elementary step of this scheme is given by
\begin{equation}
\label{swimp}
 T_4(\delta \tau) = T_P \left(\frac{\delta \tau}{6} \right) \;
                    T_U \left(\frac{\delta \tau}{2} \right) \;
                    T_P \left(\frac{2}{3} \delta \tau \right) \;
                    T_U \left(\frac{\delta \tau}{2} \right) \;
                    T_P \left(\frac{\delta \tau}{6} \right) \;\; .
\end{equation}
Note that in an elementary step of this scheme, the variation of the action 
with respect to the gauge-fields has to be computed twice.  Comparing the 
efficiency of this scheme and the leap-frog we have to keep this factor of two
in mind.

In our study we are primarily interested in the effect of the 
pseudo-fermions on the step-size of the integration scheme.
Therefore
we have employed the split of the time-scale proposed by 
Sexton and Weingarten \cite{SeWe}. This, in practice,  allows to eliminate 
completely the effect of the gauge action on the step-size.

The update of the momenta is split into two 
parts: One with respect to the gauge action and one with respect to the 
pseudo-fermion action:
\begin{eqnarray}
\label{splitleapfrog}
T_{PG} (\delta \tau) \; &:&  P \rightarrow P
- i \delta \tau \; 
\delta_U S_g(U) \\
T_{PF} (\delta \tau) \; &:&  P \rightarrow P
 - i \delta \tau \; 
\delta_U  S_f(U) \;\;.
\end{eqnarray}
The leap-frog scheme is now generalised to 
\begin{equation}
T_2(n,\delta \tau) \;=\; T_{PF} \left( \frac{\delta \tau}{2} \right) 
\left[
T_{PG} \left(\frac{ \delta \tau}{2 n} \right)
T_U   \left(\frac{ \delta \tau}{n} \right)
T_{PG} \left(\frac{ \delta \tau}{2 n} \right)
\right]^n 
 T_{PF} \left( \frac{\delta \tau}{2} \right) \;\;\;.
\end{equation}

The improved scheme in eq.~(\ref{swimp}) is generalised to 
\begin{equation}
\label{splitsw}
T_4(n,\delta \tau) \;=\; T_{PF} \left( \frac{\delta \tau}{6} \right) \;
 X\left( \frac{\delta \tau}{2} \right) \;
 T_{PF} \left( \frac{2}{3} \delta \tau \right) \;
 X\left( \frac{\delta \tau}{2} \right) \;
 T_{PF} \left( \frac{\delta \tau}{6} \right)  \;\;\;,
\end{equation}
where
\begin{equation}
 X(\delta \tau) =
 T_{PG} \left( \frac{\delta \tau}{6} \right) \;
 T_U \left(\frac{ \delta \tau}{2} \right) \;
 T_{PG} \left( \frac{2}{3} \delta \tau \right) \;
 T_U \left(\frac{ \delta \tau}{2} \right) \;
 T_{PG} \left( \frac{\delta \tau}{6} \right) \;\;\;.
\end{equation}

In our study we have used $n=4$ throughout. We have tested that larger
values of $n$ hardly further increase the acceptance rates.

\section{The Numerical Study}
We have tested our modified algorithm at parameters that had been 
studied by UKQCD before \cite{Sroczynski}. We have performed simulations
at $\beta=5.2$ and $c_{sw} = 1.76$. Note that $c_{sw} = 1.76$ was a 
preliminary result for the improvement coefficient, while the final
analysis resulted in $c_{sw} = 2.0171$ for $\beta=5.2$ \cite{alphacsw3}.
We applied periodic boundary conditions in all lattice directions,
except for anti-periodic boundary conditions in time-direction 
for the fermion-fields.

As a first step we have studied  a 
$8^3 \times 24$ lattice at $\kappa=0.137$.
Later we simulated a $16^3 \times 24$ lattice at
$\kappa=0.139$, $0.1395$ and $0.1398$. Following the tables
4.4 and 4.6 of ref. \cite{Sroczynski}, these values of $\kappa$ 
correspond to $m_{\pi}/m_{\rho} \approx 0.856, 0.792, 0.715$ and $0.686$, 
respectively.
From table 4.8 of ref. \cite{Sroczynski} we read that
$\kappa_c \approx 0.1405$ with an error of about one in 
the last digit.

\subsection{Details of the implementation}
We have simulated the two modified actions with independent programs. The 
program for the action $\tilde{S}_F$ in eq.~(\ref{rainer}) was 
written in TAO and was executed
on a APE100 computer. 
The program for the action $S_F$ in 
eq.~(\ref{orginal}) was written in C with sequences
of assembler code to take advantage of the SSE2-instructions of the Pentium 4 CPU
(See ref. \cite{Luscher2001}). Since the implementation in the two cases differ
in many respects, we give separate discussions of the details below.

\subsubsection{The action $S_F$}
Here we have used double-precision floating point arithmetic throughout.
We have implemented the BiCGstab algorithm \cite{BiCGstab} to solve the linear
equations (\ref{heat2},\ref{XYgeneral1},\ref{XYorginal2})
and to compute the action at the end of the trajectory.
 
As start vector we have always used the zero-vector.
The square of the residual is given by
\begin{equation}
\label{residual}
r^2  = |M \eta^{(m)} - \phi|^2    \;\;,
\end{equation}
where $\eta^{(m)}$ is the approximation of the solution after the $m^{th}$ 
iteration. 
Note that we did not compute $r^2$ using eq.~(\ref{residual}) directly, 
but used instead the iterated result of the BiCGstab algorithm.
The iteration is stopped after $r^2$ becomes smaller than a given 
bound.
For the heat bath of eq.~(\ref{heat2}) 
of the field $\phi_2$
and to compute the action at the end of the trajectory we have 
required $r^2 < 10^{-20}$.

As long as the start vector does not depend on the history of the 
trajectory, 
the correctness of the HMC algorithm does not  rely
on the accuracy of the vectors $X$ and $Y$ of section 
\ref{trajsect}. However, if the accuracy becomes too low, $\Delta H$ 
becomes large and hence the acceptance rate low.  

The vectors $X$ and $Y$ for $S_{F1}$ we computed as follows: 
\begin{equation}
\label{firststep}
 \tilde M Y = \gamma_5 \phi_1 \;\;,
\end{equation}
where we required 
\begin{equation}
\label{stopping}
r^2 < 0.01 R^2 
\label{defofR}
\end{equation}
as stopping criterion, where 
$R^2$ is a certain cut-off, the choices of which are listed below.
For the solution $X$ of 
\begin{equation}
\label{secondstep}
 \tilde M X = \gamma_5 Y
\end{equation}
we required $r^2 < R^2$. 
Here
we followed the recommendation of ref. \cite{Sroczynski}.
Since the error of $Y$ propagates to $X$, we can not reach the same 
accuracy of $X$ as of $Y$. Therefore it is reasonable to require 
a higher accuracy for eq.~(\ref{firststep}) than for eq.~(\ref{secondstep}).

In a similar fashion we compute the vectors $X$ and $Y$ for the second
part of the pseudo-fermion action $S_{F2}$:
\begin{equation}
\hat M Y = \rho \phi_2 \;\;,
\end{equation}
where we require $r^2 < 0.01 R^2$.
The vector $X$ is then obtained through
\begin{equation}
\tilde M X = \gamma_5 (\phi_2 + Y) \;\;,
\end{equation}
with the stopping criterion $r^2 < R^2$.

Unfortunately,
there is also a subtle
effect of the accuracy of $X$ and $Y$ on the reversibility of the algorithm.
For a discussion see refs. \cite{JLQCD,UKQCD2000,liapunov,FrezJan}
and refs. quoted therein.
We performed a few checks to investigate this problem. We find that
the  amplitude
of the reversibility violations is much the same for the leap-frog
as for the Sexton-Weingarten improved scheme. Also it does not depend much
on the parameter $\rho$ of the modified pseudo-fermion action.
On the other hand we see, in accordance with the literature, that the
reversibility violations become much worse when we compute $X$ and $Y$
with reduced accuracy.
Going a trajectory of length $1$ forward and backward again, we see for
$\kappa=0.1398$  violations in the Hamiltonian that are smaller
than  $10^{-8}$ for $R^2 = 10^{-20}$. On the other hand, for $R^2 = 10^{-4}$
the violations become as large as $10^{-2}$.
Nevertheless, these reversibility violations are
still small enough, not to invalidate the results reported below.

In the case that the BiCGstab algorithm fails to converge,
which never happened in our 
study, $X$ and $Y$ have to be computed with the conjugate gradient 
algorithm: \\
For $S_{F1}$ we 
first compute $X$ as the solution of 
\begin{equation}
 W^2 X = \phi_1
\end{equation}
followed by
\begin{equation}
 Y = W X  \;\;\;.
\end{equation}
For $S_{F2}$ we
first compute $X$ as the solution of
\begin{equation}
 \hat Q^2 X = ( \hat Q + \rho \gamma_5 ) \phi_2
\end{equation}
followed by
\begin{equation}
 Y =  \hat Q  X - \phi_2 \;\;.
\end{equation}

\subsubsection{The action $\tilde{S}_F$}
These simulations were performed with single precision arithmetic throughout.
As solver we have used the conjugate gradient algorithm. As start vector we
have used the solution of the previous step. 
The vectors $X$ and $Y$ for $\tilde{S}_{F1}$ are computed as follows: 
First we solve the 
linear equation
\begin{equation}
 [\hat Q^2 + \rho^2] X = \phi_1
\end{equation}
followed by
\begin{equation}
 Y = \tilde W^{\dag} X \;\;.
\end{equation}
For $\tilde S_{F2}$ we  solve
\begin{equation}
 \hat Q^2 X = \rho^2 \phi_1
\end{equation}
followed by
\begin{equation}
Y = \rho^{-2} \tilde W^{\dag} X \;\;.
\end{equation}
For $X$ we tried to reach essentially the machine precision. Thus, 
as stopping criterion we required that the square of the iterated residual 
divided by the square of the norm of $X$  
is smaller than $10^{-8}$. 

\subsection{Analysing the Performance of the Algorithm} \label{choice}
The goal of the optimisation of an algorithm is to obtain a  given
statistical error for the observable one wants to compute with a minimal
amount of CPU time. To this end, the ALPHA-collaboration has studied in 
a benchmark of various algorithms  \cite{alphabench} the quantity
\begin{equation}
 M_{cost} = \frac{\mbox{CPU-time}}{(\mbox{stat. error})^2} \;\;.
\end{equation}
The practical problem with this definition is that it requires  
rather large statistics to obtain reliable results for the 
integrated auto-correlation time $\tau_{int}$
and correspondingly for the statistical error.
On the other hand it requires only a rather small number (say 100) 
of trajectories to obtain an accurate estimate of the acceptance
rate. 
Therefore we shall base our study on the hypothesis that for a fixed 
length of the trajectory and a fixed acceptance rate, the 
auto-correlation times are independent on the parameter $\rho$ of the 
modified pseudo-fermion action and the integration scheme that is used.
In fact, this hypothesis is backed up by the simulations of the 
Schwinger model in ref. \cite{MH_schwinger}
and the simulations  of the $8^3 \times 24$ lattice presented below.

Following this hypothesis, we are looking for the parameter
$\rho$ that allows for the largest step-size (i.e. minimal CPU-cost)
at a given acceptance rate. This requires however fine tuning of 
$\delta \tau$ for each parameter of $\rho$. Instead we have kept 
the step-size $\delta \tau$ fixed and have searched for the value of $\rho$ 
with the maximal acceptance rate.
The number of 
iterations needed to solve the linear equations, 
given in eq.~(\ref{XYgeneral1}), depends
on $\rho$.
On the other hand the linear equations, given in eq.~(\ref{XYgeneral2}), always 
involve the matrix $\hat Q$ and hence the number of iterations
does not depend on $\rho$.
Since the number of iterations 
for eq.~(\ref{XYgeneral2}) is, at least for  $\kappa \rightarrow \kappa_c$,
much larger than that for equations eq.~(\ref{XYgeneral1}) we   
shall ignore the dependence of the total iteration number on
$\rho$ and concentrate on the acceptance rate.

\subsection{Simulation results} 

\subsubsection{Simulations of the $8^3 \times 24$ lattice}

\begin{figure}
\begin{center}
\includegraphics[width=14cm]{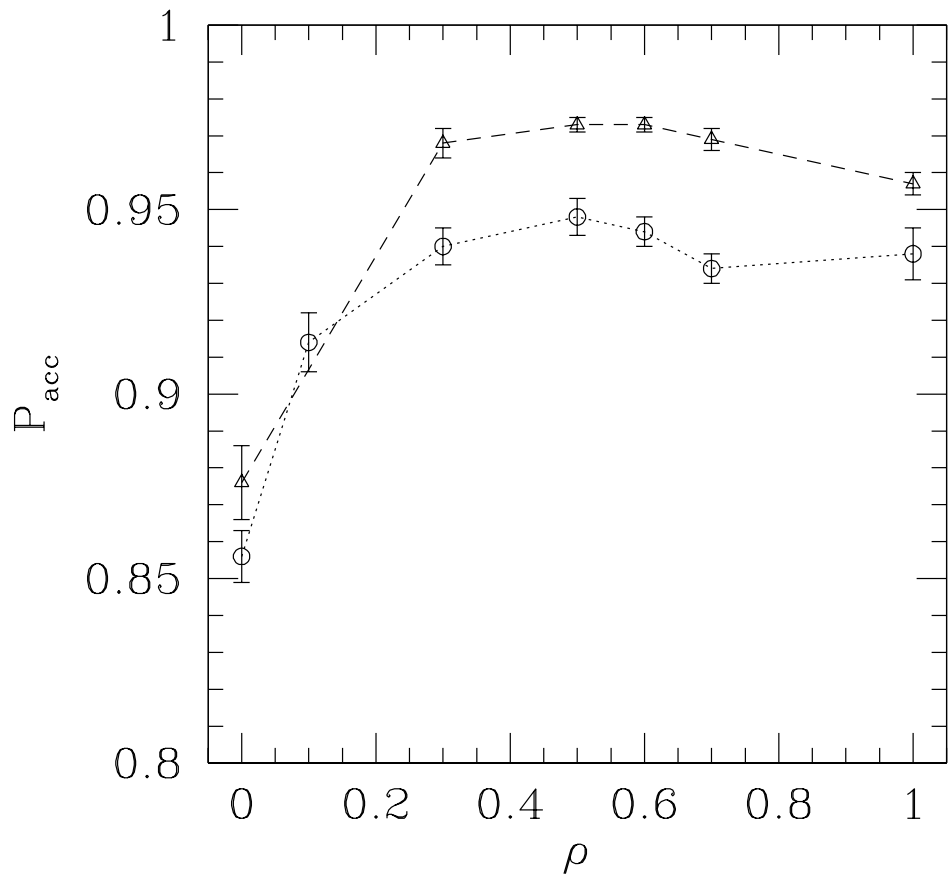}
\caption[acc]
{\label{figacc}  Results for the acceptance rate of the leap-frog scheme
(circle) and the Sexton-Weingarten improved scheme (triangle) as function 
of the parameter $\rho$. The dashed and the dotted lines should only 
guide the eye.
For the leap-frog scheme, we have fixed the 
step-size to $\delta \tau=0.02$ and for the Sexton-Weingarten improved scheme 
to $\delta \tau=0.05$. The simulations are performed on a $8^3\times 24$ 
lattice at $\beta=5.2$, $c_{sw}=1.76$  and $\kappa=0.137$. The 
pseudo-fermion action $S_F$ of eq.~(\ref{orginal}) is used.
}
\end{center}
\end{figure}

As a first step, 
we have studied quite extensively the $8^3 \times 24$ lattice at 
$\kappa=0.137$.
First we searched for the optimal value of $\rho$ for the action 
$S_F$ in eq.~(\ref{orginal}). After thermalising the system,
we performed runs with 200 trajectories each
for a range of values for $\rho$. We have fixed the step size
$\delta \tau=0.02$ for the leap-frog scheme and $\delta \tau=0.05$ for 
the  Sexton-Weingarten improved scheme. The results for the 
acceptance rate are summarised in figure  \ref{figacc}. 
In order to determine the acceptance rate, we have
averaged $\mbox{min}[1,\exp(-\Delta H)]$ instead of counting the accepted
configurations. In this way the statistical error of the acceptance rate
is considerably reduced.

We first observe that 
the acceptance rate with the modified 
pseudo-fermion action is indeed higher than
for the standard action ($\rho=0$).
For both integration schemes we see a rather broad maximum 
of the acceptance rates and the  
maximal acceptance is reached for $\rho \approx 0.5$. 

In order to compare the efficiency of the various approaches, 
we have performed more extended runs for $\rho=0$ and $\rho=0.5$ 
with the action $S_F$, eq.~(\ref{orginal}), and for 
the action $\tilde{S}_F$, eq.~(\ref{rainer}). 
In the case of $\tilde{S}_F$
we have chosen $\rho$ following the rule given in eq.~(\ref{ruleforrho}).
In these extended runs we have chosen the step-size $\delta \tau$ such that 
$P_{acc} \approx 0.8$. The parameters of these runs and the acceptance rates
are summarised in table \ref{longruns8}.  In the case of the 
leap-frog scheme the step-size can be increased by a factor of about $1.6$ by
using the modified pseudo-fermion action $S_F$,  eq.~(\ref{orginal}).
In the case of the Sexton-Weingarten improved scheme, we see a similar 
increase of the step-size by a factor of approximately $1.7$. 
The pseudo-fermion action $S_F$ seems  to perform
slightly better than the action $\tilde{S}_F$.
Comparing the leap-frog and the Sexton-Weingarten improved scheme, 
we see a small advantage for the Sexton-Weingarten improved scheme, independent
of the pseudo-fermion action that is used.

In table \ref{longruns8more} we give our results for the plaquette and the 
minimal and maximal eigenvalue of $\hat Q^2$ for the runs with the action 
$S_F$ of 
eq.~(\ref{orginal}).
The observed consistency of the results for these observables
among the three runs gives us confidence 
that the modified pseudo-fermion action has been correctly implemented in 
the program.
Also the result $P=0.49405(34)$ for the plaquette given 
in ref. \cite{Sroczynski}
is in reasonable agreement with ours. From the run with the action 
$\tilde{S}_F$ we get 
$\lambda_{min} = 0.00584(4)$, 
which is also consistent with the results obtained 
with the action $S_F$.

For the runs with the action $S_F$, eq.~(\ref{orginal}),
 we have computed integrated
auto-correlation times for the plaquette and for $\lambda_{min}$. 
We have truncated the summation of the auto-correlation function at $t=80$.
Within error-bars the integrated autocorrelation times are 
the same for the three runs. I.e. our hypothesis that
the autocorrelation times do not depend 
very much on the integration scheme and the form of the pseudo-fermion 
action, as long as the acceptance rate is the same, is confirmed.

In table \ref{longruns8solve} we give the iteration numbers
that are needed by the BiCGstab solver in the simulations with the action 
${S}_F$.
We will denote by $N_1$ the number of iterations for the action 
$S_{F1}$ ($\tilde{S}_{F1}$) and with $N_2$ the 
number of iterations for the action $S_{F2}$ ($\tilde{S}_{F2}$).
$N_1$ constitutes the 
numerical overhead caused by the extra part $S_{F1}$ ($\tilde{S}_{F1}$)
of the modified 
pseudo-fermion action. 
As can be seen in table \ref{longruns8solve}, $N_2$ depends only 
little on $\rho$, as expected, since here the original matrix 
$\hat{Q}^2$ has to be inverted.
Since most of the CPU-time is spent for the solver,  
the total number of solver-iterations per trajectory 
is a good indicator for the numerical effort. 

We see that, despite the numerical
overhead due to the additional pseudo-fermion field, a net advantage for
the modified pseudo-fermion action remains, as demonstrated by the 
total number of iterations given in the last column in 
table~\ref{longruns8solve}. As we shall see below,
this total number of 
iterations can be further reduced by choosing a larger value of $R^2$,
see eq.~(\ref{defofR}). However, 
this affects the simulation of the standard pseudo-fermion action and 
the modified pseudo-fermion action in the same way. 
Hence our conclusion on the relative performance gain is not affected.

In the simulation of the pseudo-fermion action  
$\tilde{S}_F$ we find 
the average iteration numbers of the conjugate gradient solver to be
$N_{2}=121.8(3)$ at $\rho=0$
and $N_{1}=27.278(3)$ and $N_{2}=122.1(3)$ for $\rho=0.348$. 
Note 
that here we have used the same stopping criterion 
for the calculation of the ``force'' as for 
the calculation of the action at the end of the trajectory. Again $N_{2}$ 
does not depend on $\rho$ and $N_{1}$ gives the overhead of the 
simulation with the modified pseudo-fermion action. 
It is difficult to compare the 
iteration numbers of the CG and the BiCGstab in our study, 
since different stopping criteria
have been used. 
Nevertheless, if we 
compare twice the iteration number of the 
BiCGstab with the iteration number of the CG, as we should,
we end up with a ratio of 
$N_{2}^{CG}/(2 \tilde N_{2}^{BiCG}) \approx 1.75$. 
Choosing more comparable stopping criteria, as was done in 
ref.~\cite{Sroczynski}, an advantage of
$40 \%$ in favour of the BiCGstab solver can be found. Note however that
the BiCGstab requires more linear algebra operations per iteration than
the CG. E.g. on the APE100 computer this means that the BiCGstab and the 
CG solver require approximately the same CPU-time.

\begin{table}
\caption{\sl \label{longruns8}
Results for extended runs for the 
$8^3 \times 24$ lattice with $\beta=5.2$, $\kappa= 0.137$ and
$c_{sw}=1.76$. We have used either the leap-frog (L) or the Sexton-Weingarten
improved (S) scheme; "stat" gives the number of trajectories that have been 
generated.  $P_{acc}$ is the acceptance rate.
 In the runs with the pseudo-fermion 
action $S_F$, eq.~(\ref{orginal}), 
we have averaged $\mbox{min}[1,\exp(-\Delta H)]$.
In the case of $\tilde{S}_F$, eq.~(\ref{rainer}),
we have just counted the accepted
configurations.
}
\begin{center}
\begin{tabular}{|c|c|c|c|c|c|}
\hline
scheme & action & $\rho$ &$\delta \tau$ & stat & $P_{acc}$  \\
\hline
L  &eq.~(\ref{orginal})&0.0\phantom{0...}& 0.025          & 6030 & 0.793(3)\\ 
L  &eq.~(\ref{orginal})&0.5\phantom{0...}& 0.04\phantom{0}& 8300 & 0.770(3) \\
S  &eq.~(\ref{orginal})&0.5\phantom{0...}& 0.1\phantom{00}& 6170 & 0.883(2) \\
\hline
S  &eq.~(\ref{rainer})&0.0\phantom{0...} & 0.06\phantom{0}  &2400 &
   0.83(1)\phantom{0}\\
S  &eq.~(\ref{rainer})&0.348...& 0.1\phantom{00}  &8000 & 0.798(9) \\ 
\hline
\end{tabular}
\end{center}
\end{table}

\begin{table}
\caption{\sl \label{longruns8more}
Observables for the runs with the action of $S_F$, eq.~(\ref{orginal}), of the 
previous table. $P$ is the value of the 
plaquette. $\lambda_{min}$ and $\lambda_{max}$ are the minimal and maximal 
eigenvalue of $\hat Q^2$.
Computing $\tau_{int}$ for the plaquette and  $\lambda_{min}$ 
we summed autocorrelation-functions up to $t=80$.
}
\begin{center}
\begin{tabular}{|c|c|c|c|c|c|c|}
\hline
scheme &$\rho$&$P$&$\tau_{int,P}$&$\lambda_{min}$&$\tau_{int,\lambda}$
& $\lambda_{max}$ 
 \\
\hline
L & 0.0 & 0.49474(29) & 25.8(6.0) & 0.00593(4) & 9.3(2.2) & 2.8377(11)\\
L & 0.5 & 0.49487(22) & 23.2(4.6) & 0.00591(3) &8.2(1.6)&2.8381(9)\phantom{0}\\
S & 0.5 & 0.49457(27) & 24.4(5.6) & 0.00596(4) & 9.6(2.2) & 2.8389(10) \\
\hline
\end{tabular}
\end{center}
\end{table}

\begin{table}
\caption{\sl \label{longruns8solve}
Iterations needed by the BiCGstab solver.
$ N_{1}^{acc}$ and $N_{2}^{acc}$ are the numbers of iterations that were
needed to compute $S_{F1}$ and $S_{F2}$ of eq.~(\ref{orginal}), respectively,
at the end of the trajectory. $N_{1}^{traj}$ and $N_{2}^{traj}$ are
the iteration numbers that are needed to compute the vectors $Y$ for $S_{F1}$ and 
$S_{F2}$ of eq.~(\ref{orginal}), respectively.
$N^{total}$ is the total number of iterations needed for one
trajectory. $\tau_N$ is the auto-correlation time of the total number of 
iterations.
}
\begin{center}
\begin{tabular}{|c|c|c|c|c|c|c|c|}
\hline
\rule[0mm]{0mm}{4mm}
scheme &$\rho$  & $N_{1}^{acc}$  & $N_2^{acc}$ & $N^{traj}_1 $  
& $  N^{traj}_2 $ & $\tau_N$ & $N^{total}$ \\
\hline
 L &0.0 &  -        & 60.4(2)  &  -        & 34.11(11) & 21.1(4.9)  & 2777(9) \\
L&0.5 &18.762(8)\phantom{0}&61.2(2)& 11.533(7) &
 34.89(9)\phantom{0} &19.4(3.9) & 2362(5) \\
 S &0.5 &18.750(13)& 60.9(2)  & 11.527(8) & 34.78(12) & 23.1(5.3)  & 1922(5) \\
\hline
\end{tabular}
\end{center}
\end{table}

Next we tested what accuracy of the solver is needed in the calculation of the 
variation of the action to maintain a high acceptance rate.  
For this purpose we have performed 
runs with 200 trajectories each 
for the action $S_F$ with $\rho=0$ and $\rho=0.5$ and 
the same step-sizes as for the runs reported in the table \ref{longruns8}
and various values for $R^2$.
Our results are summarised in 
table  \ref{accuracy}.  We observe that the dependence of the acceptance rate
on $R^2$ is much the same for $\rho=0$ and for $\rho=0.5$ as well as for the 
two choices of the integration scheme. Going from $R^2 = 0.1$ to $R^2 = 1$, the 
acceptance rate drastically drops. On the other hand, 
for $R^2 = 0.01 $ we have essentially  reached the acceptance rate
of $R^2 = 10^{-8}$ that is given in table \ref{longruns8}.

\begin{table}
\caption{\sl \label{accuracy}
Runs with 200 trajectories each for the $8^3 \times 24$ lattice, $\beta=5.2$, 
$c_{sw}=1.76$ and $\kappa=0.137$. The pseudo-fermion action of eq.~(\ref{orginal})
is used.  Here we study the dependence of the acceptance rate $P_{acc}$ on the 
stopping criterion $R^2$ of the BiCGstab solver, see eq.~(\ref{stopping}).
}
\begin{center}
\begin{tabular}{|c|c|c|c|c|c|}
\hline
scheme & $\rho$  & $\delta \tau$ & $R^2$  &  $P_{acc}$  &$N^{total}$ \\
\hline
 L & 0.0  & 0.025       &  1\phantom{.00} &  0.410(27) &   1022(6) \\
 L & 0.0  & 0.025       &  0.1\phantom{0} &  0.723(20) &   1230(5) \\
 L & 0.0  & 0.025       &  0.01   &  0.809(17) &   1430(6) \\
\hline
 L & 0.5  & 0.04\phantom{0}&  1\phantom{.00}&  0.390(26) &\phantom{0}941(5)\\
 L & 0.5  & 0.04\phantom{0}&  0.1\phantom{0}&  0.687(22) &   1078(5) \\
 L & 0.5  & 0.04\phantom{0}&  0.01   &  0.741(18) &   1269(6)\\
\hline
 S & 0.5  & 0.1\phantom{00} &  1\phantom{.00}&0.361(26) &\phantom{0}782(3)\\
 S & 0.5  & 0.1\phantom{00} &  0.1\phantom{0}&  0.747(19)&\phantom{0}899(5)\\
 S & 0.5  & 0.1\phantom{00} &  0.01   &  0.863(11) &  1030(4)\\
\hline
\end{tabular}
\end{center}
\end{table}

\subsubsection{Simulations of the $16^3 \times 24$ lattice}

In order to have a more difficult situation,
we went on to 
the $16^3 \times 24$ lattice with $\kappa$-values that are
closer to $\kappa_c$. In particular, 
we have performed simulations at $\kappa=0.1390$, $0.1395$ and $0.1398$
with the pseudo-fermion action $S_F$ of eq.~(\ref{orginal}). For comparison, 
we have also simulated the action 
$\tilde{S}_F$ of eq.~(\ref{rainer}) at $\kappa=0.1395$.

Our results for the acceptance rate and the total number of 
iterations of the BiCGstab solver per trajectory 
for $\kappa=0.1390$ and $0.1398$
are summarised in table \ref{many}. For each set of parameters we have
generated 100 trajectories, starting from an equilibrated configuration.
Since the integrated auto-correlation time of the acceptance rate is very
small, we can give meaningful error-estimates for this quantity. On the other
hand, as we have learned from the longer runs of the $8^3\times 24$ lattice,
the integrated auto-correlation time of the total number of iterations
is much larger. Hence we do not quote error-bars for $N^{total}$ and
consider 
$N^{total}$ to give only an indication for the relative CPU-costs of the 
simulations.

Let us start the discussion by 
testing the effect of the  stopping criterion $R^2$ of the 
BiCGstab solver on the acceptance rate.
For the leap-frog scheme  with
$\rho=0.5$ and $\kappa=0.1390$ the acceptance rate for $R^2=0.01$
is almost the same as for $R^2=0.001$, while it drastically drops as we 
further  relax the stopping criterion to $R^2=0.1$ and $R^2=1.0$. In order 
to keep the effect of $R^2$ on the acceptance rate small we have chosen
$R^2=0.001$ or $R^2=0.0001$ in the following simulations for $\kappa=0.139$.
In the case of $\kappa=0.1398$ we see a similar dependence of the acceptance 
on $R^2$. Here we have chosen $R^2=0.0001$ for most of the simulations.

\begin{figure}
\begin{center}
\includegraphics[width=14cm]{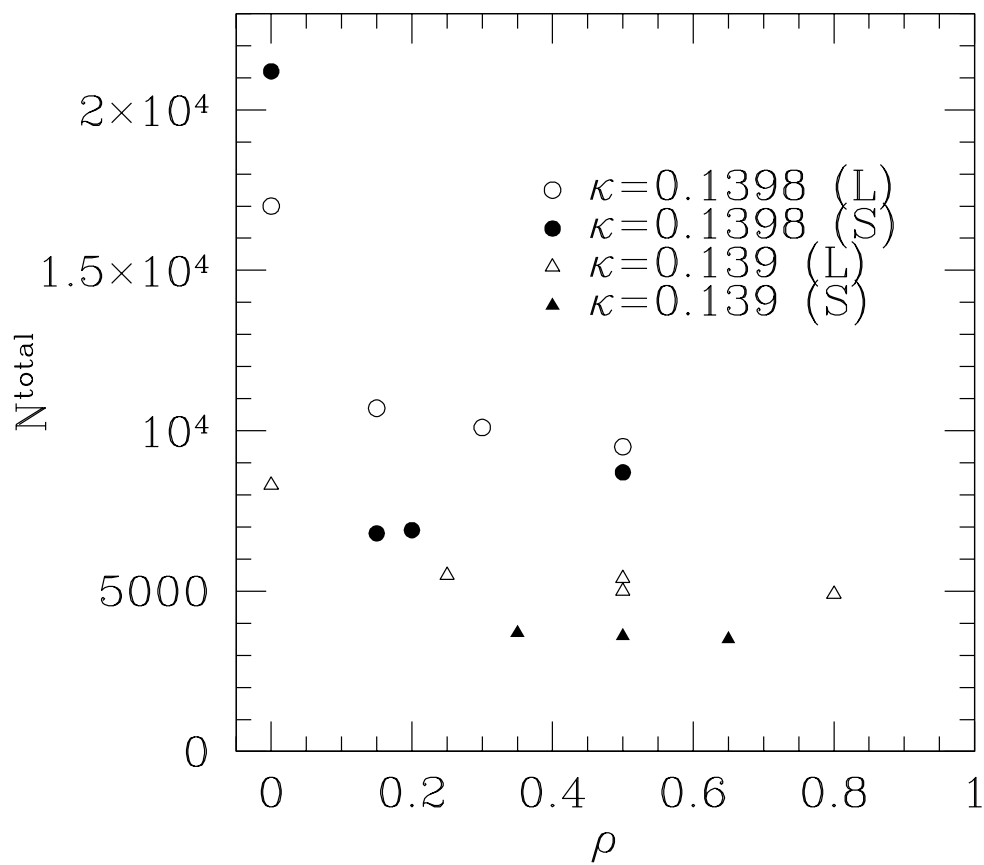}
\caption[perf]
{\label{perf}  Performance results in terms of the total number
of iterations $N^{total}$ in the BiCGstab solver as a function of
$\rho$. We denote by (L) the leap frog and by (S) the improved 
integration scheme. 
}
\end{center}
\end{figure}

Next, we studied the dependence of the acceptance rate on the parameter
$\rho$ of the modified pseudo-fermion action.
For the leap-frog scheme with $\rho \ne 0$ we have used the step-size
$\delta \tau=0.02$. 
As can be seen in table~\ref{many},
the maximum of the acceptance rate is reached at $\rho=0.5$ 
for $\kappa=0.1390$ as well as $\kappa=0.1398$.
As for the $8^3 \times 24$ lattice at $\kappa=0.137$, this maximum is broad. 
This means that
no fine tuning is needed to obtain almost optimal performance.
In order to obtain the same acceptance rate for $\rho=0$ as for $\rho=0.5$
the step-size has to be halved. 
The performance of the algorithm in terms of $N^{total}$ using
$S_F$ is shown in fig.~\ref{perf}. We selected points from
table~\ref{many} that have an acceptance rate of about 80\% .
For the case of the leap-frog integrator we find a gain
in performance by a factor of $1.66$ at $\kappa=0.1390$, which
becomes
$1.79$ at $\kappa=0.1398$ compared with the standard pseudo-fermion
action.

It is interesting to note that in contrast to our discussion in section
\ref{choice} the optimal value of $\rho$ is the same for $\kappa=0.1390$
and $\kappa=0.1398$. One explanation might be that the low eigenmodes of
the fermion matrix become relevant for the dynamics of the HMC algorithm
with the leap-frog scheme only for $\kappa$ even closer to $\kappa_c$.

For the case of 
the Sexton-Weingarten improved scheme,
we find, at $\kappa=0.1398$,
a gain of a factor of $2.4$ in favour of the modified pseudo-fermion action.
It is also interesting to note from table~\ref{many} that in this case
the optimal
value of $\rho$ decreases as $\kappa$ approaches $\kappa_c$, following 
our expectations.
The difference in the behaviour of the optimal values 
of $\rho$ between the two integration schemes is not too 
surprising. For the standard pseudo-fermion action the step-size
of higher order integration schemes has to be reduced more drastically
than for the leap-frog scheme (see e.g. ref. \cite{Takaishi}).

For $\rho=0$ at $\kappa=0.1398$ we obtain an 
acceptance rate of
$82 \%$ with the Sexton-Weingarten improved scheme and $\delta \tau=0.0166...\;$.
This means that for the standard pseudo-fermion action, 
the  Sexton-Weingarten improved scheme performs worse than the leap-frog 
scheme. On the other hand, the gain due to the modified pseudo-fermion action
is much larger for the Sexton-Weingarten improved scheme than for the 
leap-frog scheme which explains the gain in performance shown in 
fig.~\ref{perf}.

In table~\ref{many1395} we give results for $\kappa=0.1395$ 
for both pseudo-fermion actions of $S_F$ and $\tilde{S}_F$.
Here we have only used the Sexton-Weingarten improved scheme. 
The acceptance rate  as well as the step-size  for 
$S_F$ are slightly larger than for $\tilde{S}_F$.
Comparing the two modified pseudo-fermion actions we
find hence a minor advantage for $S_F$. 
Clearly, more studies have to be performed 
to be able to make a solid statement, comparing the two modified actions.

In table \ref{physres} we give results for the plaquette and for the 
maximal and minimal eigenvalues of $\hat Q^2$. The numbers that are quoted are 
naive averages over all runs listed in table \ref{many} and the
runs with the action $S_F$ in table \ref{many1395}. 
The error-bars
are naively computed from the fluctuation of the averages of the single runs.
Since the computation of the smallest and the largest eigenvalue of 
$\hat Q^2$ is rather CPU-consuming, we had measured it for $\kappa=0.1390$ and 
$\kappa=0.1398$ only once at the beginning of the run. For $\kappa=0.1395$ we 
measured the eigenvalues after each trajectory. Our results for the plaquette
for $\kappa=0.1390$ and   $\kappa=0.1398$ are consistent within the 
quoted error-bars with those of \cite{Sroczynski}. In the case of 
$\kappa=0.1395$ there is a $2 \sigma$ discrepancy. The average
value of $\lambda_{max}$ is, within error-bars, identical for all three 
values of $\kappa$ that we had studied. On the other hand, $\lambda_{min}$
is decreasing as $\kappa_c$ is approached. Fitting the three values with 
the ansatz
\begin{equation}
\lambda_{min} = C (\kappa_c - \kappa)^x
\end{equation}
we obtain 
$x=1.85 \pm 0.09$, where we have fixed $\kappa_c=0.1405$.

\begin{table}
\caption{\sl \label{many}
 Results for the $16^3 \times 24 $ lattice at $\beta=5.2$ and
 $c_{sw}=1.76$ with the pseudo-fermion action $S_F$ of eq.~(\ref{orginal}).
 For each set of parameters we have generated 100 trajectories, starting 
 from an equilibrated configuration.
}
\begin{center}
\begin{tabular}{|l|c|l|l|l|l|r|}
\hline
  \multicolumn{1}{|c}{$\kappa$}
& \multicolumn{1}{|c}{scheme}
& \multicolumn{1}{|c}{$\rho$}
& \multicolumn{1}{|c}{$R^2$}
& \multicolumn{1}{|c}{$\delta \tau$ }
& \multicolumn{1}{|c}{$P_{acc}$} 
& \multicolumn{1}{|c|}{$N^{total}$} \\
\hline
 0.139 &  L    &  0.   & 0.001& 0.01 &  0.82(2) &  8300\\
 0.139 &  L    &  0.05 & 0.001& 0.02 &  0.69(3) &  6600\\  
 0.139 &  L    &  0.1  & 0.001& 0.02 &  0.75(3) &  6200\\
 0.139 &  L    &  0.25 & 0.001& 0.02 &  0.78(3) &  5400\\
 0.139 &  L    &  0.5  & 1.   & 0.02 &  0.12(3) &  3200\\
 0.139 &  L    &  0.5  & 0.1  & 0.02 &  0.68(4) &  3700\\
 0.139 &  L    &  0.5  & 0.01 & 0.02 &  0.77(2) &  4500\\ 
 0.139 &  L    &  0.5  & 0.001& 0.02 &  0.81(2) &  5000\\
 0.139 &  L    &  0.8  & 0.001& 0.02 &  0.77(2) &  4900\\
 0.139 &  L    &  1.2  & 0.001& 0.02 &  0.69(3) &  4100\\
 0.139 &  S    &  0.1  & 0.0001& 0.066...&0.61(3) & 4400\\ 
 0.139 &  S    &  0.2  & 0.0001& 0.066...&0.75(3) & 3970\\
 0.139 &  S    &  0.35 & 0.0001& 0.066...&0.78(2) & 3700\\
 0.139 &  S    &  0.5  & 0.0001& 0.066...&  0.80(3)& 3600\\
 0.139 &  S    &  0.65 & 0.0001& 0.066...&  0.76(2)& 3500\\ 
 0.139 &  S    &  0.8  & 0.0001& 0.066...&  0.68(3)& 3400\\
 0.139 &  S    &  1.0  & 0.0001& 0.066...&  0.60(3)& 3500\\
\hline
 0.1398&  L    &  0.   & 0.0001&0.01 &  0.77(3) & 17000\\
 0.1398&  L    &  0.05 & 0.0001&0.02 &  0.62(3) & 11700\\
 0.1398&  L    &  0.15 & 0.0001&0.02 &  0.75(3) & 10700\\
 0.1398&  L    &  0.3  & 0.0001&0.02 &  0.76(2) & 10100\\
 0.1398&  L    &  0.5  & 0.1   &0.02 & 0.53(4)  & 5800\\
 0.1398&  L    &  0.5  & 0.01  &0.02   &0.75(3) & 7000\\
 0.1398&  L    &  0.5  & 0.0001&0.02 &  0.78(2) & 9500\\
 0.1398&  L    &  1.0  & 0.0001&0.02 & 0.64(4)  & 9300\\
 0.1398&  S    & 0.    & 0.0001&0.02 & 0.64(4)  &16500\\
 0.1398&  S   & 0.    &0.0001& 0.0166...& 0.82(2)&21200 \\
 0.1398&  S & 0.05 & 0.0001& 0.066... & 0.35(4)  & 7500 \\
 0.1398&  S & 0.1  & 0.01& 0.066...   & 0.46(4)  & 5100 \\
 0.1398&  S & 0.1  & 0.0001& 0.066... & 0.67(3)  & 6500\\
 0.1398&  S & 0.15 & 0.0001& 0.066... & 0.72(3)  & 6800\\
 0.1398&  S & 0.2  & 0.01  & 0.066... & 0.60(3)  & 4700\\
 0.1398&  S & 0.2  & 0.0001& 0.066... & 0.74(3)  & 6900\\
 0.1398&  S & 0.4  & 0.0001& 0.066... & 0.49(3)  & 6100\\
 0.1398&  S & 0.5  & 0.0001& 0.04545 ...& 0.82(2) & 8700 \\
\hline
\end{tabular}
\end{center}
\end{table}

\begin{table}
\caption{\sl \label{many1395}
 Results for the $16^3 \times 24$ lattice at $\kappa=0.1395$. For the 
 action of eq.~(\ref{orginal}) we have chosen $R^2=0.001$ as stopping
 criterion of the BiCGstab solver. In all cases, we have used the 
 Sexton-Weingarten improved scheme. Each of the runs with the 
  action of eq.~(\ref{orginal}) consists of 100 trajectories  while the 
 run with the action  of eq.~(\ref{rainer}) consists of 380 trajectories.
 We have used a step-size of $\delta \tau=0.057$ and  $17$ 
 steps per trajectory. I.e. the trajectory length is not exactly equal to 1.
 In the case of the action eq.~(\ref{orginal}), 
 we computed the acceptance rate $P_{acc}$ 
 by  averaging $\mbox{min}[1,\exp(-\Delta H)]$,
 while for the action of 
 eq.~(\ref{rainer}) we just have counted the accepted configurations.
 $N_1$ and $N_2$ give the number of iterations needed by the solvers 
 for the computation of the force due to $S_{F1}$ and $S_{F2}$, 
 respectively. Note that the BiCGstab has to be applied twice, while
 the CG is only applied once to compute the force.
}
\begin{center}
\begin{tabular}{|c|l|l|l|r|r|}
\hline
 action &\phantom{o} $\rho$ &\phantom{o} $\delta \tau$&\phantom{o}$P_{acc}$ &
 $ N_1 $ \phantom{o} & $N_2$ \\
\hline
 \ref{orginal} &  0.05 & 0.066...& 0.40(3) &30.2\phantom{0}&  70 \\ 
 \ref{orginal} &  0.2  & 0.066...& 0.78(3) &15.4\phantom{0}&  67 \\
 \ref{orginal} &  0.5  & 0.066...& 0.66(3) & 9.2\phantom{0}&  61 \\
\hline
 \ref{rainer}  &  0.2236 & 0.057  & 0.75(3) &31.25& 223 \\
\hline
\end{tabular}
\end{center}
\end{table}

\begin{table}
\caption{\sl \label{physres}
Results for the observables from the simulations of the
$16^3 \times 24 $ lattice. For comparison we give the result for plaquette
obtained in ref. \cite{Sroczynski} ($P_{UKQCD}$). A discussion of the 
error-bars is given in the text. 
}
\begin{center}
\begin{tabular}{|c|c|c|c|c|}
\hline
$\kappa$ & $P_{UKQCD}$  & $P$  & $\lambda_{max}$ & $\lambda_{min}$ \\
\hline
 0.1390 &0.51593(14)& 0.51602(12) & 2.95(2)\phantom{0}& 0.00110(4)\phantom{0}\\
 0.1395 &0.52196(9)\phantom{0}& 0.52174(9)\phantom{0}& 2.960(4)  & 0.000513(7) \\
 0.1398 &0.52477(12)& 0.52466(11) & 2.95(2)\phantom{0}& 0.00028(2)\phantom{0}\\
\hline
\end{tabular}
\end{center}
\end{table}

\section{Conclusion and outlook}
In ref. \cite{MH_schwinger} we suggested that the Hybrid-Monte-Carlo
simulation of dynamical Wilson fermions can be substantially speeded
up by a modification of the pseudo-fermion action. The basic idea
is that the fermion matrix is split up into two factors
 such that each factor has a
smaller condition number than the original fermion matrix. 
For each of the
factors, a pseudo-fermion field is introduced. In  ref. \cite{MH_schwinger}
we found a speed-up of more than
a factor of two for the largest value of $\kappa$
that we simulated.

In the present paper we have developed the idea of ref. \cite{MH_schwinger}
in a more general fashion. In addition to the original factorisation we
discuss a second that is inspired \cite{Sommerprivate} 
by twisted mass QCD \cite{twisted}.
Note that in our approach only one additional
free parameter $\rho$ appears as compared with the standard HMC simulations.
Our numerical results show that no fine tuning of the parameter 
$\rho$ is needed.
We argued that the condition number for the modified fermions is 
changed to the square root of the original condition numbers. This
bears the potential of changing the scaling behaviour of the 
HMC algorithm when the chiral limit is approached.
We have also demonstrated that the modified pseudo-fermion action can be
easily used on top of even-odd preconditioning of the clover-improved
Wilson fermion matrix.

In our numerical study we find that for lattice-spacings, quark masses and
lattice sizes that recently have been used in large scale simulations
\cite{Sroczynski} a reduction of the numerical cost of more
than a factor of two compared with the standard pseudo-fermion action 
can be achieved.

One interesting observation in our study 
is that the partially improved scheme of Sexton and Weingarten does 
profit more from the modification of the pseudo-fermion action than the 
leap-frog scheme. In fact, this is not too surprising, since higher 
order schemes have more problems in the limit $\kappa \rightarrow \kappa_c$ 
than the leap-frog scheme. This is nicely studied in \cite{Takaishi} and also
found in recent large scale simulations \cite{CP-PACS}. Based on our 
experience it would be interesting to study genuinely higher order 
integration schemes combined with the modified pseudo-fermion action.

There are a number of directions in which the present work could be extended.
Our general framework~(\ref{general}) allows for a large variety of 
factorisations of the fermion matrix beyond our two choices of eqs.~(\ref{orginal}) 
and~(\ref{rainer}).
For example, one could divide the lattice in sub-lattices
and construct the matrix $W$ by eliminating all hopping terms 
from $\hat Q$ that connect different sub-lattices. Such a construction might 
be particularly useful for a massively parallel computer.

An obvious idea is to enlarge the number of factors  in eq.~(\ref{general}).
This might be particularly advantageous, when we go to light quark masses.
In order to keep the condition number of the factors of the fermion matrix
constant as the quark mass becomes lighter, more factors have to be 
introduced.

A related modification of the HMC algorithm has been proposed and tested in 
refs. \cite{PeardonLattice,PeardonLattice2002}. The first difference compared
with our approach is the factorisation of the fermion matrix:
\begin{equation}
\tilde M = P(M)^{-1} \;\;,
\end{equation} 
where $P(M)$ is a low order polynomial approximation of $M^{-1}$.  
The order of the polynomial plays a similar role as the parameter
$\rho$ in our approach. The use of the polynomial might allow for 
a better separation of the UV-part of the spectrum than our factorisation.
On the other hand, the polynomial $P(M)$ requires the introduction 
of an auxiliary field for each order of the polynomial. 

In contrast to us, the authors of ref. 
\cite{PeardonLattice,PeardonLattice2002} put the two contributions of 
the pseudo-fermion action on different time scales of the integration 
scheme. In our case it would also be possible to put $S_{F1}$ on the 
same time scale as $S_G$.  In fact, preparing ref. \cite{MH_schwinger} 
we have tested this idea. However we found no advantage compared with 
our present setting. In order to keep the algorithm as simple as possible, 
we did not discuss this possibility here. 

An interesting idea is to apply the idea of refs. 
\cite{PeardonLattice,PeardonLattice2002} to the 
Polynomial-Hybrid-Monte-Carlo algorithm \cite{PHMC1,PHMC2}. I.e. to use a 
polynomial approximation for both parts of the pseudo-fermion action:
\begin{equation}
S_{F1} = |P_1(M) \phi_1|^2\;\;, \;
S_{F2} = |P_2(M) \phi_2|^2\;\;,
\end{equation}
where now $P_1(M)$ is a rough, low order approximation of $M^{-1}$ and 
the product $P_1(M)  P_2(M)$ is an accurate approximation of $M^{-1}$. 
This idea is particularly 
appealing since it could well be applied to the simulation 
of three flavour QCD. (See ref. \cite{JLQCD} and refs. therein).

\section{Acknowledgements}
We thank  R. Sommer for discussions. We are grateful to M. L\"uscher who 
provided us with his benchmark code for the Dirac-operator.
This work was supported in part by the European Community's Improving
Human Potential Programme under contracts HPRN-CT-2000-00145
(Hadrons/Lattice QCD) and HPRN-CT-2002-00311 (EURIDICE).

\end{document}